\documentclass[english,11pt]{article}

\usepackage{amssymb}
\usepackage{graphicx}
\usepackage{amsfonts}
\usepackage{bbm}
\usepackage[T1]{fontenc}
\usepackage[latin1]{inputenc}
\usepackage{babel}
\usepackage{color}

\newcommand{\be}{\begin{equation}}

\newcommand{\ee}{\end{equation}}

\newcommand{\rf}[1]{(\ref{eq:#1})}

\begin{document}

\title{On  the   Stability of  Black Holes  at the  LHC}

\author{M. D. Maia\thanks{maia@unb.br}\\Universidade de Brasilia, Instituto de Física, Brasilia, 70910-970 \\ \& \\
E. M. Monte\thanks{Edmundo@fisica.ufpb.br}\\
Universidade Federal da Paraíba,  Departamento de Física, 8059-970}

\maketitle

\begin{abstract}
The  eventual    production  of  mini  black holes  by  proton-proton collisions  at the  LHC   is  predicted by
theories with   large extra  dimensions  resolvable  at   the Tev  scale of  energies.  It is  expected  that these   black holes evaporate shortly  after  its  production  as  a  consequence of the Hawking radiation.
We  show  that  for  theories    based on the  ADS/CFT correspondence,    the  produced  black holes  may have an  unstable  horizon,  which grows proportionally  to  the  square of  the  distance  to the collision point.
\end{abstract}

The  production of  mini black holes  at  the LHC  Tev  energy  scale is predicted   by  theories   with  large  extra  dimensions and   where the  gravitational  constant is  replaced  by  a  Tev   scale fundamental  constant  \cite{Dimopoulos}.  In those theories the gauge interactions   are  confined to  four-dimensional  space-times   embedded  in  a  bulk space defined by the Einstein-Hilbert  principle,    but gravitation  as  described by  the  four-dimensional  embedded geometry,  can  access   the extra  dimensions at  the  TeV   energy. Thus, the  generated  black  hole will  represent  a  local deformation of  Minkowski's  space-time   resulting  from the  proton-proton collision.
Typical  examples  are  given by the    Randall-Sundrum  models    of  brane-world theory  defined in    the 5-dimensional  anti deSitter $AdS_5$  bulk,  using   the  Israel-Lanczos jump  condition   at   the  brane-world,   acting as    a boundary  \cite{RS}.

The  produced   black holes   are  expected  to be   short  lived as  consequence of  Hawking's  radiation.  The  well known  theorem  by  Hawking  of  1975, made use of  semi-classical   gravitation   to prove that  virtual particle pairs are formed in the vicinity of a  black hole. While one  of the particles of the pair falls inside the black hole horizon,   the other escapes to  infinity   producing a  thermal  radiation  of  the black hole. Admitting that the    radiation is  complete the black hole will evaporate  away,  with  the    consequence is  that  the correlation between the spin states of the particles pair  is lost, compromising the  quantum unitarity  \cite{Hawking}. In  a  recent (2005) revision  of  his  theorem, Hawking   made use of   the ADS/CFT correspondence   within the context of the Horava-Witten  $AdS_5 \times  S^5$  model of  string theory.   It was  assumed that the  black holes  in question are  charged  (like for example, the Reissner-Nordstrom black hole), so that after the evaporation  a  stable submanifold  would remain,  in which case some of  the  information  would be  stored   and  could be  partially retrieved \cite{Hawking:1}.

In  any  of the  above  mentioned models  using  the same $AdS_5$   space,  the  stability  of  the    black  hole,    seen  as  a   subspace  embedded    in  that  bulk  must  be understood.

\vspace{4mm}

Consider the four-dimensional space-time $ AdS_4$,   regarded  as  a   hypersurface  with negative constant curvature   embedded in the five dimensional flat space $M_5 (3,2)$, with maximal isometry $SO(3,2)$ (see  eg \cite{Rosen,Eisenhart}). Adding  to  each of these  spaces  one extra spatial dimension,  we obtain  in a very trivial way   that  the   five-dimensional anti de Sitter space $AdS_5$ is  a  hypersurface  with negative constant curvature  embedded in the  flat space  $ M_6 (4,2)$, with maximal isometry $SO(4,2)$,  which  is    isomorphic to the  conformal group $C_0$ acting on the  Minkowski space-time, so that  for   each  conformal covariant  Yang-Mills field  in   $M_4$, there  corresponds   an  isometric  covariant Yang-Mills field  defined  in the  $AdS_5$   space.

Since  all  Yang-Mills    fields  and  their  duality  properties  are  consistently  defined only  in four-dimensional  space-times,  where   3-forms  are  isomorphic  to  1-forms,  the  ADS/CFT   correspondences  must  be  hold  between  four-dimensional  subspaces  of  the  $AdS_5$  bulk. Consequently,    the  isometric  invariant  Yang-Mills fields  must  remain  confined  to  a  4-dimensional  subspace  embedded  in the    $AdS_5$  bulk,   where   the  required  duality operation  is preserved.  Furthermore, admitting that  the embedding of the  4-dimensional
subspace is local  and   regular, then   the   inverse  function theorem    establishes    a  local   1:1 correspondence  between    the  conformal  fields  in  $M_4$  and the  Yang-Mills   fields defined in those   four-dimensional   subspaces.

The   $AdS_5$ bulk is  a  solution of  the 5-dimensional vacuum  Einstein's  equations  with negative  cosmological  constant
\begin{equation}
^5{\cal R}_{AB}-\frac{1}{2}\, ^5{\cal R} {\cal G}_{AB}  -\Lambda   {\cal G}_{AB}  =0,  \;\;\;  A,B=1...5 \label{eq:bulkEE}
\end{equation}
and  with constant  curvature,    characterized by  the  Riemann tensor
\be
 ^5{\cal R}_{ABCD}= \frac{\Lambda}{6}({\cal G}_{AC}{\cal
G}_{BD}-{\cal G}_{AD}{\cal G}_{BC})
 \label{eq:bulk}
\ee
The embedding  is  an  application   $X: V_4 \rightarrow AdS_5$,  with components   $X^A (x_\mu)$,  functions of the   space-time coordinates $x_\mu$.  Together  with  the  unit normal  vector  $\eta$,  this  defines  a Gaussian  reference frame   on  the  embedded  geometry  $\{ X^A_{,\mu},  \eta^A  \}$.  The  Riemann tensor  of the bulk   written in this  reference  gives the integrability  equations for the embedding,  the  well known
 Gauss-Codazzi  equations  \cite{Eisenhart},  which in  the  case of  a constant  curvature  5-dimensional bulk can be  written as
\begin{eqnarray}
&&\hspace{-7mm}R_{\alpha\beta\gamma\delta} =
 -(k_{\alpha\gamma}k_{\beta\delta}-
k_{\alpha\delta}k_{\beta\gamma}) +
\frac{\Lambda}{6} (g_{\alpha\gamma}g_{\beta\delta}-
g_{\alpha\delta}g_{\beta\gamma})\label{eq:G1}\\
&&\hspace{-7mm}k_{\alpha[\beta;\gamma]} = 0, \label{eq:C1}
\end{eqnarray}
The  existence of  the  embedding  requires that the     metric $g_{\mu\nu}$ and  the extrinsic  curvature $k_{\mu\nu}$ of  $V_4$   satisfy  these  equations.   Writing   \rf{bulkEE} in the  same Gaussian  frame    we obtain the  gravitational equations  of the  embedded $V_4$ \cite{Maia:1}
\begin{eqnarray}
&&\label{eq:Einstein2}R_{\alpha\beta}-\frac{1}{2}Rg_{\alpha\beta}-
\Lambda g_{\alpha\beta} + {Q}_{\alpha\beta}  = 0\\
&&k^\gamma_{\beta;\gamma}-h_{,\beta}=0
\end{eqnarray}
where  $h=g^{\mu\nu}k_{\mu\nu}$  is the mean  curvature,  $K^2 =  k_{\mu\nu}k^{\mu\nu}$  is the Gaussian  curvature and
$$Q_{\alpha\beta}  = ( k^{\rho}{}_{\alpha}k_{\rho\beta
}-h k_{\alpha\beta}) -\frac{1}{2}(K^{2}-h^{2}) \,\;g_{\alpha\beta}, \;\;\; Q^{\alpha\beta}{}_{;\beta} =0  $$
These  equations  describe the  dynamics of the  gravitational  field  for  a  space-time  embedded  in the  $AdS_5$ bulk.
Clearly  they are   more general  than the vacuum  Einstein's  equations in general relativity,  because   \rf{bulkEE}
implies  that   the   extrinsic   curvature  $k_{\mu\nu}$  is  also  a  dynamical  variable.  When  $k_{\mu\nu}=0$   the usual vacuum Einstein's  equations  are  recovered.

Now, we  may    discuss the  production of  black hole  at the LHC
with   the  supposition that  it  happens  within  the  $AdS_5$
bulk.
The  experiment  is  supposed to take place in  the   Minkowski's  space-time $M_4$  where   the protons  are    defined.
However, this $M_4$ is     regarded  as  a  subspace  embedded in the  $AdS_5$,    so that    equations  \rf{G1} and  \rf{C1}  must  apply.  The  general   solution  of these equation for  a  flat  space gives   $k_{\mu\nu}= \sqrt{{\Lambda}/{6}}\;\eta_{\mu\nu}$.    This means
that   although  $M_4$  is   flat in  the Riemann sense, as an embedded geometry it  is also warped  like  a cylinder, a  cone or  any  ruled surface,   where  the   full  translational  group  of the   Poincaré  symmetry  in principle  would not  apply. However,  from \rf{Einstein2},  it   follows that    $\Lambda$  is the  cosmological  constant, which   is  too  small  to  mark a significant presence in the  local gravitational  field of the LHC  experiment. Therefore,  for  any practical purposes    we   may   assume  that the  experiment  starts very  approximately  as planned,   in Minkowski's  space-time.

 After the collision, the   produced black-holes  will   represent a  deformations of  the original Minkowski's  space-time,  transforming into Schwarzschild or   Reissner-Nordstrom     subspaces  embedded  in the   $AdS_5$ bulk. This black hole  should  remain  for  a very  short period,   before its  eventual  evaporation.
For a     spherically symmetric diagonal metric, the  general  solution of  \rf{C1}  is of the  form  $k_{\mu\nu}=  \alpha(r) g_{\mu\nu}$,  where     $\alpha(r)  =1/y(r)$  and  where  $y(r)$ is the local  center of  curvature of the  embedded geometry, which is   a value of the  extra  coordinate satisfying the  condition   $\det(g_{\mu\nu} -y(r) k_{\mu\nu})=0$ \cite{Eisenhart}.
 Replacing  this  expression in   \rf{Einstein2}, we  obtain  the   Schwarzschild-anti-de Sitter  solution
\begin{eqnarray*}
&&\phantom{x}\hspace{-8mm} ds^2\!\! =\!\! (1\!\! -\! \frac{2m}{r}\!+(
3\alpha(r)^2\!\!-\!\!\Lambda) r^2)^{-1}dr^2 +r^2 d\omega^2
\hspace{3mm}-(1\!\!-\!\!\frac{2m}{r} + ( 3\alpha(r)^2\!\!-\!\!\Lambda )
r^{2})dt^2
\end{eqnarray*}
Repeating the same  for the  Reissner-Nordstrom solution, we obtain the   Reissner-Nordstrom-anti-de Sitter  solution
\begin{eqnarray*}
&&\phantom{x}\hspace{-8mm} ds^2\!\! =\!\! (1\!\! -\!
\frac{2m}{r} \!+\!\frac{q}{r^2}\!+\!( 3\alpha(r)^2\!\!-\!\!\Lambda
) r^2)^{-1}dr^2 \!\!+r^2 d\omega^2\!\!
-(1\!\!-\!\!\frac{2m}{r}\!+\!\frac{q}{r^2} +\! (
3\alpha(r)^2\!\!-\!\!\Lambda )
r^{2})dt^2
\end{eqnarray*}
In both cases  we  obtain a black  holes whose  horizon   grows
indefinitely  with  $r^2$, thus  creating an  unstable  situation
unless we  impose the  additional  condition that  $3\alpha (r)^2$  is  constant   and  equal  to $\Lambda$. However,   this  implies that
the  black hole is  composed of umbilicus points (all of  its  directions  look like  the  same  principal  directions),  which is  not the
case  of  a   black  hole.   Even  if  we  neglect  the local influence of  $\Lambda$  as   before the collision,   we cannot  neglect  $\alpha(r)$  on  account  of the   characteristics  of  a black hole geometry.

We  conclude that  the  exterior  gravitational  field of  a Black  hole  is  not    native of  an  $AdS_5$ bulk  and  that the  black holes produced by   proton-proton collision at  the LHC  may  be unstable.   Nonetheless,  it  is possible that  in  a  higher  dimensional  bulk $D>5$,  the  behavior  of the   black  holes  is  stable.  This  follows  from the   well known  example   given by    the 6-dimensional  flat  bulk     $M_6 (4,2)$, whose  metric is  also invariant  under   $SO(4,2)$,  so that it has   the same   group  of isometries of the  $AdS_5$.  Consequently, all arguments  of the  ADS/CFT correspondence which   depend only on the Lie group properties,  can  be  extended  without  loss  of  generality to that   flat bulk. By the same argument used in the ADS/CFT  correspondence, the  quantum unitarity of the Yang-Mills  fields is  maintained  in  the six-dimensional flat bulk.

\end{document}